\documentclass[aps,prr,superscriptaddress,twocolumn,floatfix]{revtex4-2}
\usepackage{graphicx}
\usepackage{dcolumn}
\usepackage{amsmath}
\usepackage{amssymb}
\usepackage{color}
\usepackage{multirow}
\usepackage{url}
\usepackage{booktabs}
\usepackage{tabularx} 

\newcommand\hly{\bgroup\markoverwith
  {\textcolor{yellow}{\rule[-.5ex]{.1pt}{2.5ex}}}\ULon}

\newcommand{\mSR}{\textmu SR}
\newcommand{\vxmSR}{vx-\textmu SR}

\newcommand{\refsubfig}[2]{\hyperref[#1]{\ref*{#1}#2}}

\usepackage[utf8]{inputenc}
\usepackage{siunitx} 
\usepackage{verbatim} 
\usepackage[normalem]{ulem}
\usepackage{placeins} 
\usepackage{float} 
\usepackage[hidelinks]{hyperref} 
\hypersetup{
colorlinks=true,    
linkcolor=blue,
citecolor=blue,
urlcolor=blue,
pdfborder={0 0 0},
bookmarksopen=true,
bookmarksopenlevel=2,
pdfpagemode=UseOutlines,
pdfstartview={Fit},
pdftitle={},
pdfauthor={Z. Salman},
pdfsubject={Timing for Vertex-Reconstructed Muon-Spin Spectroscopy},
pdfkeywords={},
pdfhighlight= /I
}

\begin{document}
\title {High-Resolution Timing for Vertex-Reconstructed Muon-Spin Spectroscopy Using Plastic Scintillators and MuTRiG}

\author{Konrad Briggl}
\email[Corresponding author: ]{kbriggl@kip.uni-heidelberg.de}
\affiliation{Kirchhoff Institute for Physics, Heidelberg University, Im Neuenheimer Feld 227, 69120 Heidelberg, Germany}
\author{Maxime Lamotte}
\email[Corresponding author: ]{maxime.lamotte@psi.ch}
\affiliation{PSI Center for Neutron and Muon Sciences, 5232 Villigen PSI, Switzerland}
\author{Marius Snella Köppel}
\affiliation{Institute for Particle Physics and Astrophysics, ETH Zürich, CH-8093 Zürich, Switzerland}
\author{Jonas A. Krieger}
\affiliation{PSI Center for Neutron and Muon Sciences, 5232 Villigen PSI, Switzerland}
\author{Heiko Augustin}
\affiliation{Physics Institute, Heidelberg University, Im Neuenheimer Feld 226, 69120 Heidelberg, Germany}
\author{Niklaus Berger}
\address{Institute for Nuclear Physics, Johannes Gutenberg-Universität Mainz, Johann-Joachim-Becher-Weg 45, 55099 Mainz, Germany}
\author{Andrin Doll}
\affiliation{PSI Center for Neutron and Muon Sciences, 5232 Villigen PSI, Switzerland}
\author{Pascal Isenring}
\affiliation{PSI Center for Neutron and Muon Sciences, 5232 Villigen PSI, Switzerland}
\affiliation{Physik Institut, University of Zürich, Winterthurerstrasse 190, CH-8057 Zürich}
\author{Hubertus Luetkens}
\affiliation{PSI Center for Neutron and Muon Sciences, 5232 Villigen PSI, Switzerland}
\author{Sebastian M\"{u}hle}
\affiliation{PSI Center for Neutron and Muon Sciences, 5232 Villigen PSI, Switzerland}
\author{Thomas Prokscha}
\affiliation{PSI Center for Neutron and Muon Sciences, 5232 Villigen PSI, Switzerland}
\author{Thomas Rudzki}
\affiliation{Physics Institute, Heidelberg University, Im Neuenheimer Feld 226, 69120 Heidelberg, Germany}
\author{André Schöning}
\affiliation{Physics Institute, Heidelberg University, Im Neuenheimer Feld 226, 69120 Heidelberg, Germany}
\author{Hans-Christian Schultz-Coulon}
\affiliation{Kirchhoff Institute for Physics, Heidelberg University, Im Neuenheimer Feld 227, 69120 Heidelberg, Germany}
\author{Zaher~Salman}
\email[Corresponding author: ]{zaher.salman@psi.ch}
\affiliation{PSI Center for Neutron and Muon Sciences, 5232 Villigen PSI, Switzerland}

\date{\today}

\begin{abstract}
Vertex-reconstructed muon-spin spectroscopy (\vxmSR) based on silicon pixel detectors has recently demonstrated unprecedented lateral resolution and operation at muon stop rates exceeding~\SI{400}{kHz}. However, the intrinsic timing resolution of current silicon pixel detector technology limits the accessible frequency range and restricts \mSR\ measurements with fast relaxation rates. In this work, we report on the integration of plastic scintillator detectors (PSD) read out with the MuTRiG ASIC into the MuSiP \vxmSR\ spectrometer. This complements the spatial resolution achieved by using silicon pixel detectors with high-precision timing information for incoming muons and decay positrons. We demonstrate stable operation of MuTRiG in vacuum and achieve sub-\SI{300}{ps} time resolution after time-walk correction. Standard transverse-field \mSR\ measurements on a SiO$_2$ sample confirm that the combined MuTRiG–PSD system resolves precession frequencies beyond \SI{50}{MHz}, far exceeding the capabilities of silicon pixel detectors alone. These results establish a viable and scalable path towards high-rate, high-resolution \mSR\ with both excellent spatial and temporal performance.
\end{abstract}
\maketitle

\section{Introduction}
Muon-spin rotation, relaxation, and resonance (\mSR) is a powerful technique for probing magnetism, superconductivity, and quantum phenomena in condensed matter systems. At continuous muon sources, conventional \mSR\ spectrometers are typically limited to stopped-muon rates of the order of \SI{40}{kHz} to avoid pile-up and excessive background~\cite{amato_introduction_2024}, and they provide no intrinsic lateral resolution on the sample. This restricts measurements on small or spatially inhomogeneous samples and leads to an inefficient use of the available muon flux.

To address these limitations, vertex-reconstructed \mSR\ (\vxmSR) using silicon pixel detectors has been developed in recent years~\cite{Mandok2026PRR,Augustin2025NIaMiPRSAASDaAE,Isenring2025}. The \mSR\ using Si-pixel detectors (MuSiP) spectrometer employs MuPix11 pixel chips~\cite{Augustin2021JPSCP}  to reconstruct the muon stopping position and positron decay vertex with high precision, enabling background suppression and lateral imaging of the sample. MuSiP can operate at incoming muon rates exceeding \SI{400}{kHz} while maintaining a low uncorrelated background~\cite{Mandok2026PRR}.

A key remaining limitation of the current MuSiP implementation is its timing resolution. The MuPix11 chip provides a time resolution of \SI{16}{ns}~\cite{rudzki_ultra-light_2023}, which is insufficient to resolve fast muon-spin precession frequencies and is significantly worse than the sub-nanosecond performance of state-of-the-art \mSR\ spectrometers based on scintillators and silicon photomultipliers (SiPMs)~\cite{Amato2017RSI,Stoykov2012PP}. Two strategies can be envisaged to overcome this limitation: (i) the development and deployment of silicon pixel sensors with substantially improved timing performance, or (ii) the addition of dedicated fast timing detectors to complement the pixel-based spatial information. Here, we pursue the second approach since suitable fast pixel chips (i.e., with a low material budget, large area, and low heat generation) are not yet available.

In this paper, we demonstrate that the integration of plastic scintillator detectors (PSDs) read out by the MuTRiG ASIC provides a realistic and effective solution for precise timing in \vxmSR. Building on detector concepts successfully employed in conventional \mSR\ spectrometers, we show that MuTRiG-based readout can be seamlessly integrated into the MuSiP data acquisition system and delivers timing performance comparable to established systems~\cite{Amato2017RSI}. Commercial alternatives to MuTRiG exist, such as Weeroc devices commercialized by CAEN; however, our choice was primarily pragmatic since the MuTRiG can be used with the existing MuSiP hardware and software infrastructure, thereby minimizing the need for additional development effort and resources. This choice does not constitute a comparative evaluation of alternative solutions.

\begin{figure*}[htb]
    \centering
    \includegraphics[width=0.65\linewidth]{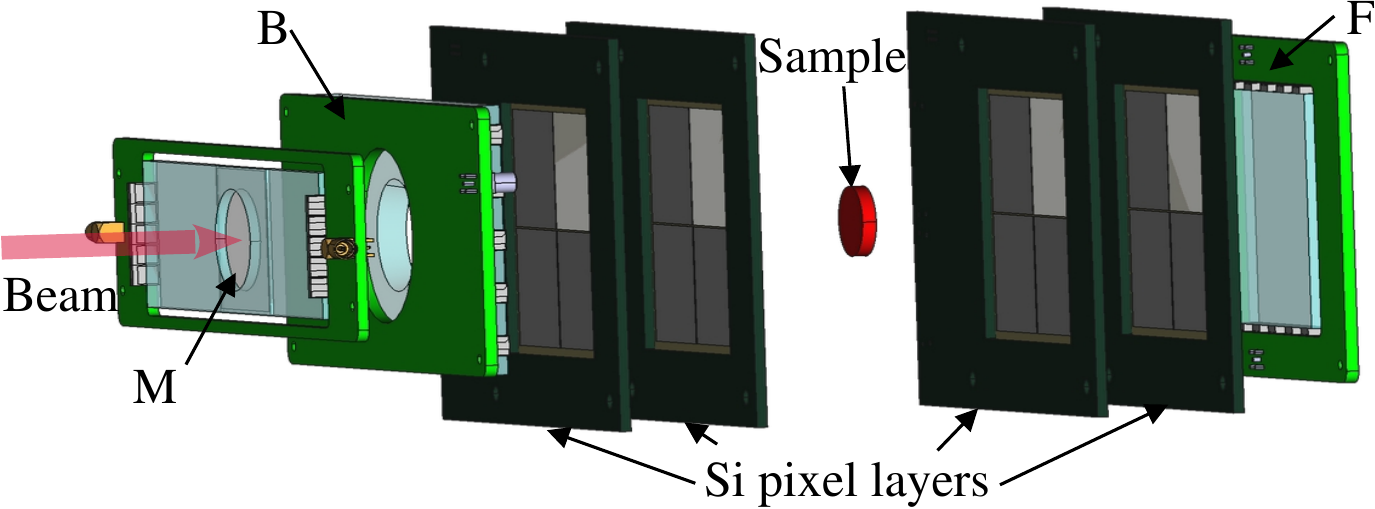}
    \caption{A schematic of the detector arrangement of the Si pixels and PSDs relative to the beam direction.}
    \label{Detectros}
\end{figure*}
\begin{figure}[bht] 
        \includegraphics[width=0.75\linewidth]{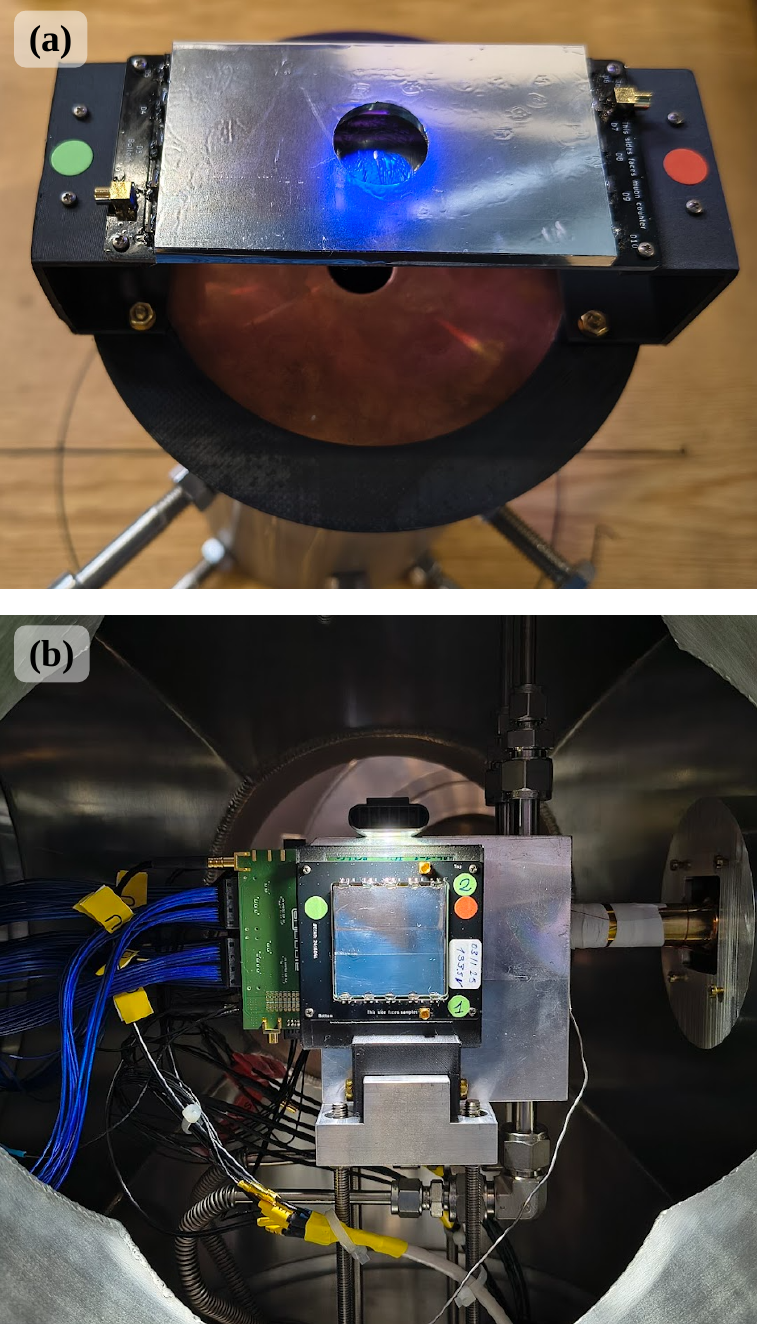}
        \caption{(a) Photograph of the copper collimator, M counter and B positron detector. (b) View of the rear side of the spectrometer. Note the F positron detector placed after the downstream MuPix11 array.}
        \label{fig:3d}
\end{figure}

Three plastic scintillators were installed in the MuSiP~\cite{Mandok2026PRR} spectrometer: a muon counter (M), a backward positron detector (B) placed upstream, and a forward positron detector (F) placed downstream, as shown in Fig.~\ref{Detectros}. All M, B, and F detectors were made of high-efficiency, fast-timing EJ-212 plastic scintillators. These were diamond polished on all faces and covered by a reflective \SI{2}{\micro m} thick aluminum-coated Polystyrene wrapping; see Fig.~\ref{fig:3d}.
The M~counter had a sensitive diameter of \SI{17}{mm} and a thickness of \SI{200}{\micro m}. The B~detector measured $85\times45\times4$~mm$^3$ and featured a central \SI{17}{mm} diameter hole to allow incoming muons to pass and reach the sample (Fig.~\ref{fig:3d}(a)), while the F~detector measured $40\times40\times4$~mm$^3$ (Fig.~\ref{fig:3d}(b)). Each scintillator was read out at both ends using two independent channels. Each channel comprised five series-connected Advansid NUV3S-P SiPMs, selected for their high sensitivity in the blue light spectrum. This redundancy allows for a possible coincidence readout from both sides of each PSD, suppressing dark counts if high-sensitivity to minimum-ionizing particles (MIPs) is desired. The SiPMs were read out AC-coupled to the MuTRiG ASIC, a 32-channel mixed-signal readout chip originally developed for the Mu3e experiment~\cite{Chen2017JI}. MuTRiG provides precise time stamping with an intrinsic resolution of the order of \SI{50}{ps} and supports hit rates of up to \SI{1}{MHz} per channel\cite{Chen2017JI}. As the readout ASIC was initially designed for single-ended or quasi-differential DC-coupled readout of the SiPMs, a special high-voltage AC-coupling adapter add-on board was designed and produced for our application.

The six SiPM channels (32-33 from M, 34-35 from B, and 36-37 from F) were connected to a MuTRiG readout PCB with \SI{30}{cm} long RG174 high-frequency cables with MMCX connectors. The readout board was installed inside the vacuum vessel and was passively cooled by thermal contact with the vessel wall. This marks the first successful operation of MuTRiG in vacuum. For the measurements reported here, the SiPM channels were operated at a nominal bias voltage of \SI{150}{V}, provided by an external low-noise high voltage power supply (PSI SCSHV20, maximum \SI{300}{V}/\SI{1}{mA} per channel), as used on most PSI \mSR\ instruments.

\section{Results}
\subsection{Scintillator signal characteristics and channel tuning}
\begin{figure}[htb]
\centering
\includegraphics[width=.75\linewidth]{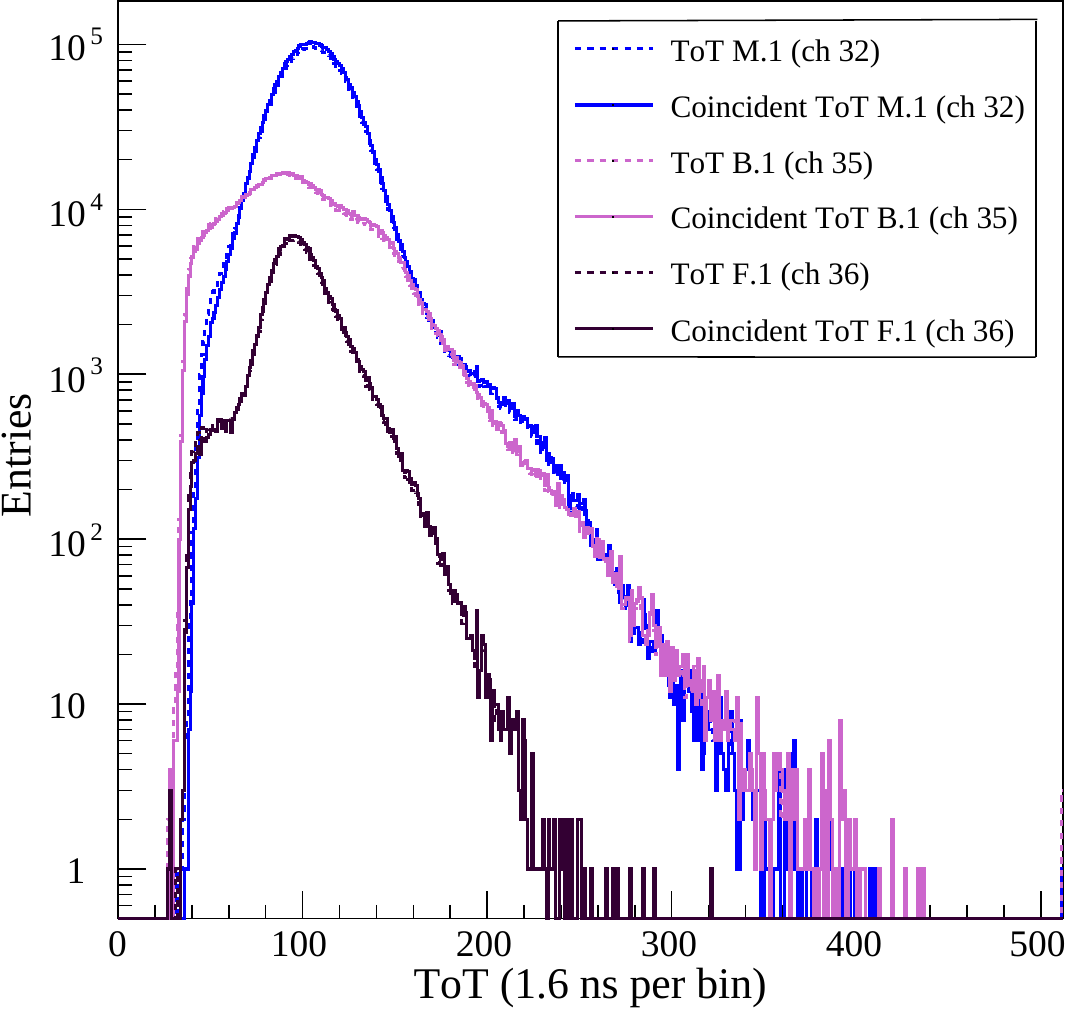}
\caption{ToT distributions from each plastic scintillator counter, shown for hits on a single readout channel without (dashed lines) and with coincidence (solid lines) with the other readout on the same PSD. The asymmetric shape observed for B is attributed to light scattering from the central hole.} \label{fig:tot_all}
\end{figure}

The timing threshold of the MuTRiG ASIC was set to its maximum value, which corresponds to a threshold of a few photo-electrons, i.e., it is much lower than the signal of a single particle. The energy threshold is used for the validation of large-amplitude signals and energy estimation using Time-over-Threshold measurements (ToT). They are comparable to the amplitude of the incoming particles and were manually tuned to retain only actual hit events. Fig.~\ref{fig:tot_all} shows representative ToT distributions from a single channel of each scintillator (dashed lines), reflecting the energy deposition of traversing particles. No significant noise hits or inefficiencies are observed when selecting only coincidence hits from the two readouts on each PSD (solid lines). The B detector exhibits a characteristic signal shape due to its central hole, which suppresses signals from incoming muons. This can be attributed to the fact that the hole scatters light emitted by particle hits, leading to a longer distribution tail on the readout side located farther from the hit position.

\begin{figure*}[htb]
    \centering
    \includegraphics[width=\linewidth]{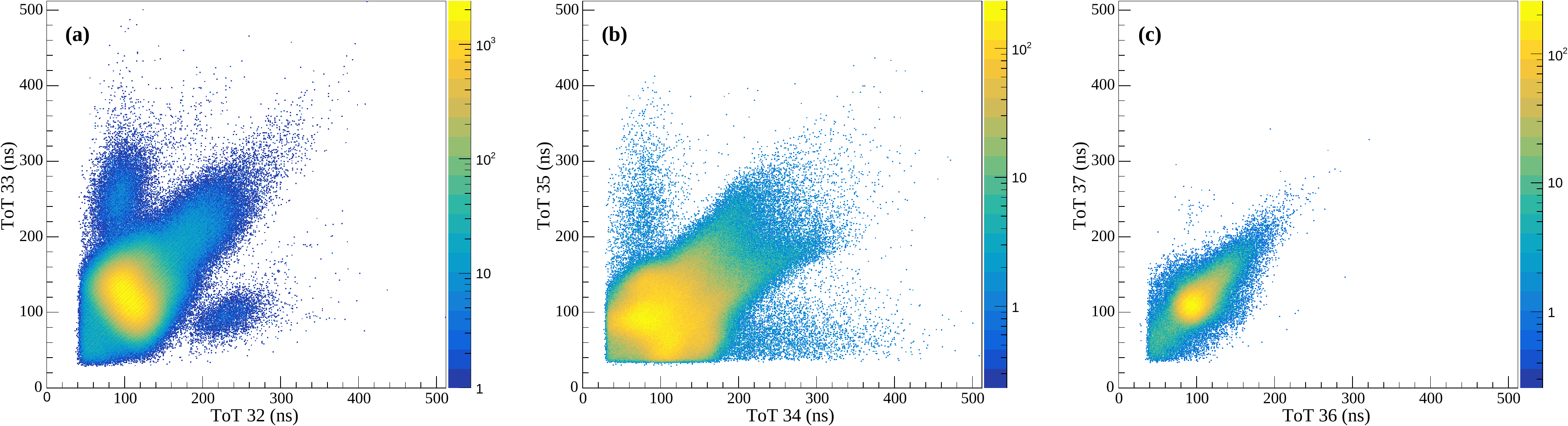}
    \caption{Correlation of the ToT measured on the two readout channels of each PSD: (a) M counter (channels 32-33), (b) B detector (channels 34-35), and (c) F detector (channels 36-37). Only hit pairs with time difference $| \Delta t |<4$~ns are shown. The clear correlations demonstrate consistent light sharing and stable SiPM response.}
    \label{fig:tot_corell}
\end{figure*}

The correlations between the ToT values measured on the two channels of each scintillator demonstrate good light distribution and a consistent response (Fig.~\ref{fig:tot_corell}). These correlations provide the basis for subsequent time-walk corrections, as described below.

\subsection{Time resolution and time-walk correction}
\begin{figure*}[bht]
    \centering
    \includegraphics[width=\linewidth]{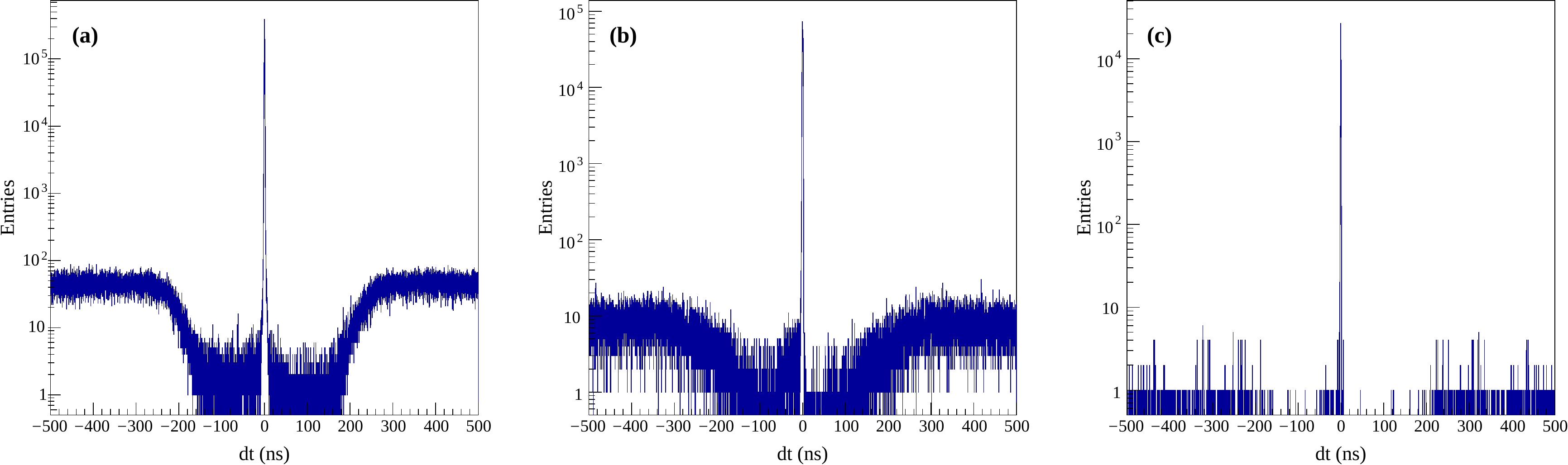}
    \caption{Raw time-difference distributions between hit pairs recorded on the two channels of the same PSD for the (a) M counter, (b) B detector, and (c) F detector.}
    \label{fig:dt_ch_wide}
\end{figure*}

Coincidence candidates were constructed from simultaneous hits originating from the same PSD block and detected on the two corresponding channels. The raw time differences, $dt$, between such hit pairs are shown in Fig.~\ref{fig:dt_ch_wide} for the M counter, B, and F detectors. Due to the use of a fixed leading-edge discriminator, the measured hit time depends on the signal amplitude: pulses with larger amplitudes cross the discriminator threshold earlier than smaller pulses. This effect, commonly referred to as time-walk \cite{Spieler2005}, leads to a systematic correlation between the measured time and the deposited energy, which in our system is encoded by the ToT value. Such behavior is well known in scintillator–SiPM readout systems and becomes particularly pronounced in beam environments with a broad distribution of particle types and energy depositions \cite{stoykov2012time,Spieler2005}.

\begin{figure}[htb]
    \centering
    \includegraphics[width=0.85\linewidth]{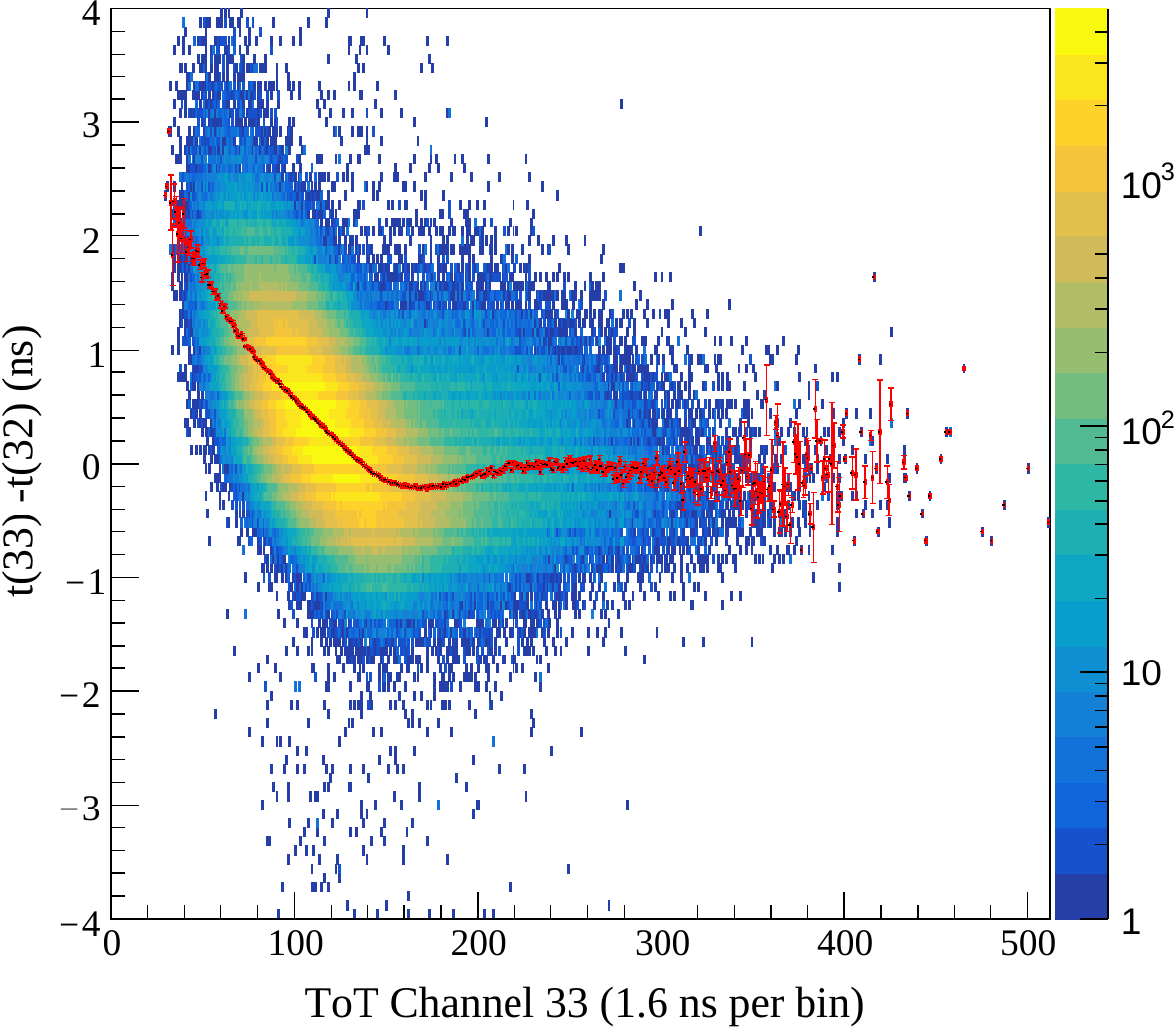}
    \caption{Time-walk distribution for the M counter. The time difference between the M counter channels 32 and 33 is shown as a function of the measured ToT. The red curve indicates the mean time offset extracted in each ToT bin and used to construct a lookup-table-based time-walk correction.}
    \label{fig:timewalk_33}
\end{figure}

\begin{figure}[bth]
    \centering
    \includegraphics[width=0.85\linewidth]{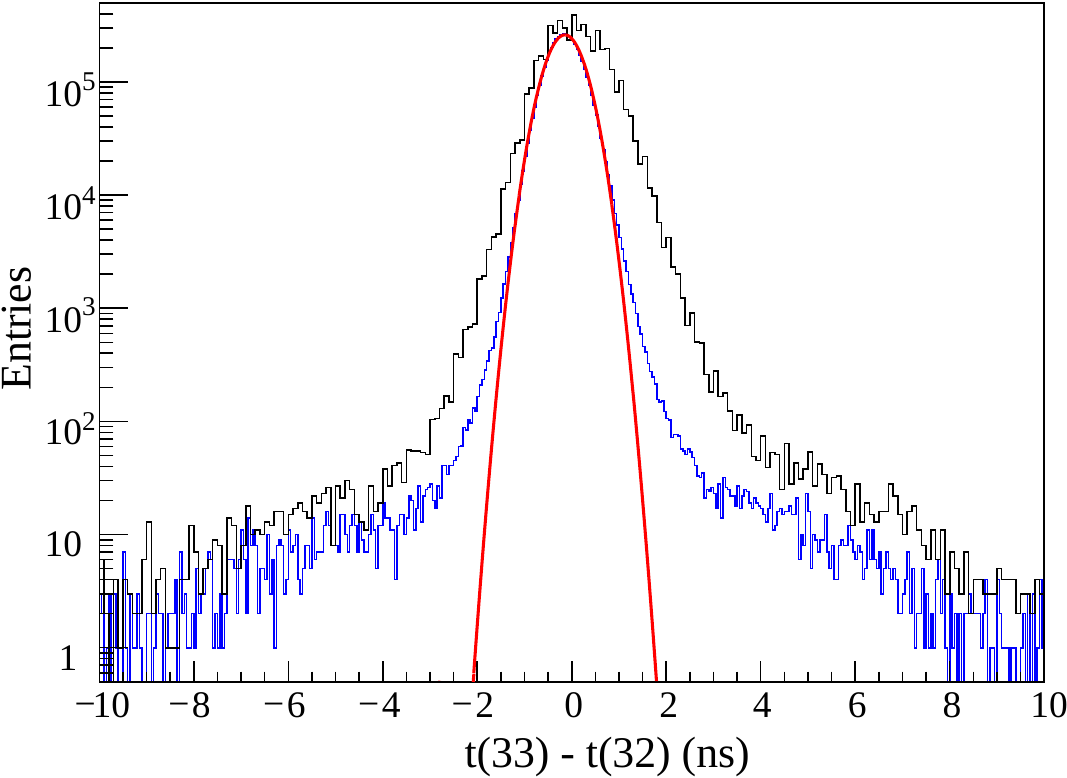}
    \caption{Time-difference distribution for the muon counter before (black) and after (blue) application of the time-walk correction derived from Fig.~\ref{fig:timewalk_33}. The red line is a fit of the time-walk corrected data to a Gaussian.}
    \label{fig:dt_corrected}
\end{figure}

To mitigate this effect, a \SI{\pm5}{ns} coincidence window was applied, and a time-walk correction was derived using a lookup-table approach, in which the mean time offset is determined as a function of ToT and subtracted on an event-by-event basis (Fig.~\ref{fig:timewalk_33}). Although this procedure, applied to each particular ASIC channel-SiPM group, does not fully capture all correlations, it significantly narrows the time-difference distributions, as shown in Fig.~\ref{fig:dt_corrected}. The time-walk correction yields a single-detector time resolution of $\sigma / \sqrt{2} =$\SI{267}{ps} for the M counter, obtained from the Gaussian fit (red curve).
A single Gaussian does not fully describe the distribution and misses the tails of the peak. This is expected since the tails originate from events with lower amplitude and thus increased time measurement jitter. Therefore, we quote the time resolution as a fit-independent metric, using the root mean square (RMS) values obtained from events in a \SI{4}{ns} window around zero, as shown in Table~\ref{tab:timing}.
\begin{table*}
    \centering
    \renewcommand{\arraystretch}{1.2}      
    \setlength{\tabcolsep}{10pt}           
    \begin{tabular}{|l|ccc|}
    \hline
         & M & B & F\\
    \hline
     MuTRiG (raw) & \SI{605}{ps} & \SI{879}{ps} & \SI{452}{ps}\\
     MuTRiG (time-walk corr.) & \SI{393}{ps} & \SI{378}{ps} & \SI{321}{ps}\\
     MuTRiG (single channel) & \SI{278}{ps} & \SI{268}{ps} & \SI{227}{ps}\\
     Offline measurement & \SI{235}{ps} & \SI{223}{ps} & \SI{195}{ps}\\
    \hline
    \end{tabular}
    \caption{Summary of time resolution obtained with MuTRiG in beam, based on the RMS of the distribution in a \SI{4}{ns} window around zero, and a reference offline measurement based on waveform sampling and analysis using a radioactive source.}
    \label{tab:timing}
\end{table*}

Within a \SI{\pm 4}{ns} window centered around zero, the RMS time difference is reduced from \SI{605}{ps}, \SI{879}{ps}, and \SI{452}{ps} to \SI{393}{ps}, \SI{378}{ps}, and \SI{321}{ps} for the M counter, B, and F scintillators, respectively. Assuming equal contributions from both channels (i.e., equal light detection efficiency), this corresponds to single-detector time resolutions of \SI{278}{ps} (M), \SI{268}{ps} (B), and \SI{227}{ps} (F). These values are consistent with those obtained in a test setup using a well-collimated $^{241}$Am \SI{5.49}{MeV} alpha source and the time resolution measurement technique described by Stoykov et al\cite{stoykov2012time}. This method relies on high-rate waveform acquisition combined with an offline implementation of a virtual constant-fraction discrimination algorithm. Using this method, time resolutions of \SI{235}{ps}, \SI{223}{ps}, and \SI{195}{ps} were obtained for M, B, and F, respectively. A comparison of the timing resolutions derived from the different analysis techniques is summarized in Table~\ref{tab:timing}. These results indicate that neither the MuTRiG ASIC nor the relatively long signal cables introduce a measurable degradation of the timing performance. Instead, the resolution achieved under beam conditions was primarily limited by time-walk effects associated with the diverse particle species incident on the scintillators. For reference, plastic scintillator detectors of comparable design have been deployed in the versatile muon spectrometer (VMS)~\cite{ToniMaxime} at PSI, which uses conventional VME-based electronics and a v1190B time-to-digital converter, yielding timing resolutions of the same order. Further gains during beam operation are expected through improved correction procedures and by incorporating tracking information from the silicon pixel detectors.

\subsection{\mSR\ performance demonstration}
Although the full integration of pixel-based spatial information with MuTRiG timing data is still ongoing, an independent data acquisition and analysis chain has been implemented for the MuTRiG-PSD system. Using this setup, standard transverse-field \mSR\ measurements were performed on a SiO$_2$ (Suprasil) sample in a low magnetic field of \SI{\sim3.5}{mT} provided by an external electromagnet. This measurement serves as a benchmark for the timing performance under realistic \mSR\ operating conditions, using a reference sample with well-known, high-frequency oscillatory components. Suprasil is commonly used for this purpose because the resulting characteristic precession frequencies of the muonium (a hydrogen-like bound state of a positive muon and electron~\cite{amato_introduction_2024}) signal provide a demanding test of sub-nanosecond time resolution, well beyond the range accessible with MuPix11 pixel-based timing alone.

\begin{figure*}[bht]
\centering   \includegraphics[width=0.95\linewidth]{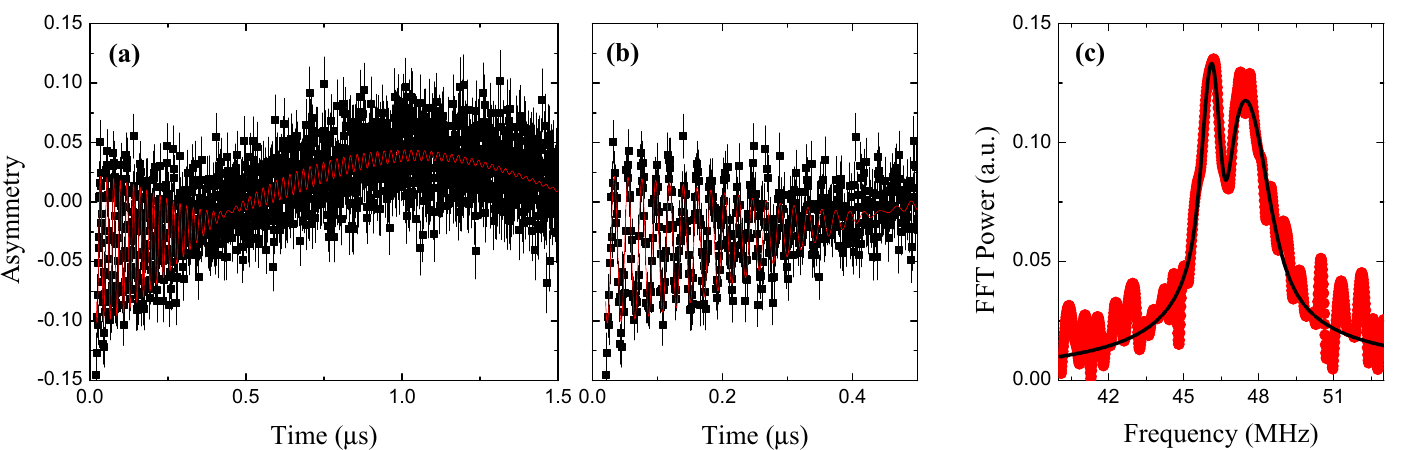}
    \caption{Transverse-field \mSR\ measurement on a SiO$_2$ (Suprasil) sample at $B\simeq3.5$~mT and room temperature using the MuTRiG-PSD system. (a) Asymmetry as a function of time over \SI{1.5}{\micro s} and (b) zoom into the first \SI{0.5}{\micro s}, illustrating the fast oscillatory signal. (c) Power spectrum obtained from the fast Fourier transform (FFT) of the asymmetry, revealing the muonium precession frequency at approximately \SI{50}{MHz}. The solid lines show the fit curve as described in the text.}
    \label{Quartz}
\end{figure*}
Fig.~\ref{Quartz} shows a representative \mSR\ asymmetry spectrum measured at \SI{\sim 3.5}{mT}, along with its Fourier transform. The measured asymmetry was fitted to the sum of three oscillating signals. The first oscillation frequency, from the diamagnetic muons fraction, corresponds to the Larmor precession in a field of $B_0=3.451(6)$~mT. The other two frequencies, \SI{46.22(2)}{MHz} and \SI{47.31(3)}{MHz} are attributed to the muonium precession frequencies in $B_0$ \cite{Brewer2000PBCM,Doll2025PRR}. The clear observation of the muonium precession frequencies near \SI{\sim 50}{MHz} demonstrates that the timing performance of the MuTRiG-PSD system is sufficient for resolving fast oscillations that are entirely inaccessible with MuPix11 alone.

\section{Discussion}
The results presented here demonstrate that adding fast plastic scintillators read out by MuTRiG effectively overcomes the principal timing limitation of the current MuSiP \vxmSR\ spectrometer. The achieved sub-\SI{300}{ps} time resolution is comparable to that of modern scintillator-based \mSR\ instruments \cite{Amato2017RSI} and represents an improvement of nearly two orders of magnitude over the intrinsic timing of the MuPix11 pixels\cite{Augustin2021JPSCP}.

Crucially, this improvement is obtained without compromising the high-rate capability and spatial resolution that define \vxmSR. This can be understood by the fact that the inaccuracy in vertex reconstruction is dominated by particle scattering in the innermost layers near the sample \cite{Mandok2026PRR,Isenring2025}. Therefore, the spatial resolution should not be affected by the addition of an outer timing layer. The PSDs provide precise global timing, while the silicon pixels retain their role in vertex reconstruction and background suppression. This division of tasks allows each detector technology to operate in its optimal regime. Note that to avoid pileup, highly segmented PSDs will be required for future applications.

The demonstrated resolution of muonium precession frequencies around \SI{50}{MHz} confirms that the combined system fully supports conventional \mSR\ modes, including transverse-field measurements in low and moderate magnetic fields. In the context of recent advances in \vxmSR\cite{Mandok2026PRR,Augustin2025NIaMiPRSAASDaAE,Isenring2025}, this development closes an important performance gap and enables experiments that require high lateral resolution and high timing precision, such as studies of small magnetic samples and other spatially inhomogeneous quantum materials.

From a technical perspective, the successful operation of MuTRiG in vacuum, its adaptability to AC-coupling of high-voltage detectors, and its seamless integration into the MuSiP data acquisition represent important milestones and further demonstrate its versatility. The measured performance suggests that further gains are achievable through optimized threshold settings, improved time-walk correction algorithms, and combined analysis with pixel-based tracking information. Scaling the system to a larger number of timing channels also appears feasible within the MuTRiG architecture.

\section{Summary and Conclusions}
We have integrated plastic scintillator timing detectors, read out by the MuTRiG ASIC, into the MuSiP \vxmSR\ spectrometer and demonstrated their performance in a beam test at the Swiss Muon Source (S$\mu$S). After time-walk correction, time resolutions of \SI{278}{ps}, \SI{268}{ps}, and \SI{227}{ps} were achieved for the M counter, B, and F detectors, respectively. Standard \mSR\ measurements on SiO$_2$ confirm that this timing performance enables the resolution of fast precession frequencies, which are well beyond the capabilities of silicon pixel detectors alone~\cite{Mandok2026PRR}.

These results establish the hybrid MuTRiG–PSD approach as a practical and powerful solution for high-precision timing in \vxmSR\ with the currently available technology. To handle the high rates of incoming muons and emitted positrons, highly segmented PSDs will be required for future applications. In addition, the segmented PSD timing layers can be used as a coarse tracking layer, providing a third layer for vertex reconstruction in magnetic fields. Together with the spatial capabilities of silicon pixel detectors, this development paves the way for next-generation \mSR\ instruments that combine high rate capability, excellent background suppression, lateral resolution, and state-of-the-art timing performance.

\section{Acknowledgments}
This research was funded by the Swiss National Science Foundation (SNSF) [\href{https://data.snf.ch/grants/grant/215167}{215167}].
All experiments were performed at the Swiss Muon Source (S\textmu S), Paul Scherrer Institute, Villigen, Switzerland.

\bibliography{references}

@article{stoykov2012time,
  title={A time resolution study with a plastic scintillator read out by a Geiger-mode avalanche photodiode},
  author={Stoykov, A and Scheuermann, R and Sedlak, K},
  journal={Nuclear Instruments and Methods in Physics Research Section A: Accelerators, Spectrometers, Detectors and Associated Equipment},
  volume={695},
  pages={202--205},
  year={2012},
  publisher={Elsevier}
}

@article{Mandok2026PRR,
  title = {Advanced Muon-Spin Spectroscopy with High Lateral Resolution Using {{Si-pixel}} Detectors},
  author = {Mandok, Lukas and Isenring, Pascal and Augustin, Heiko and Berger, Niklaus and K{\"o}ppel, Marius and Krieger, Jonas A. and Luetkens, Hubertus and Prokscha, Thomas and Rudzki, Thomas and Sch{\"o}ning, Andr{\'e} and Salman, Zaher},
  year = 2026,
  month = jan,
  journal = {Physical Review Research},
  volume = {8},
  number = {1},
  pages = {013092},
  publisher = {American Physical Society},
  doi = {10.1103/f4jb-39yn}
}

@article{Augustin2025NIaMiPRSAASDaAE,
    title = {New {Frontiers} in muon-spin spectroscopy using {Si}-pixel detectors},
    volume = {1080},
    issn = {0168-9002},
    url = {https://www.sciencedirect.com/science/article/pii/S0168900225004826},
    doi = {10.1016/j.nima.2025.170681},
    journal = {Nuclear Instruments and Methods in Physics Research Section A: Accelerators, Spectrometers, Detectors and Associated Equipment},
    author = {Augustin, Heiko and Berger, Niklaus and Doll, Andrin and Isenring, Pascal and Köppel, Marius and Krieger, Jonas A. and Luetkens, Hubertus and Mandok, Lukas and Prokscha, Thomas and Rudzki, Thomas and Schöning, André and Salman, Zaher},
    year = {2025},
    pages = {170681},
}

@misc{Isenring2025,
    title = {The effect of magnetic fields on vertex reconstructed muon-spin spectroscopy},
    url = {http://arxiv.org/abs/2510.10094},
    doi = {10.48550/arXiv.2510.10094},
    publisher = {arXiv},
    author = {Isenring, Pascal and Salman, Zaher},
    month = oct,
    year = {2025},
    note = {arXiv:2510.10094 [physics]},
}

@article{Amato2017RSI,
    title = {The new versatile general purpose surface-muon instrument ({GPS}) based on silicon photomultipliers for $\mu${SR} measurements on a continuous-wave beam},
    volume = {88},
    issn = {0034-6748},
    url = {https://aip.scitation.org/doi/10.1063/1.4986045},
    doi = {10.1063/1.4986045},
    number = {9},
    journal = {Review of Scientific Instruments},
    publisher = {American Institute of Physics},
    author = {Amato, A. and Luetkens, H. and Sedlak, K. and Stoykov, A. and Scheuermann, R. and Elender, M. and Raselli, A. and Graf, D.},
    month = sep,
    year = {2017},
    pages = {093301},
}

@article{Chen2017JI,
    title = {{MuTRiG}: a mixed signal {Silicon} {Photomultiplier} readout {ASIC} with high timing resolution and gigabit data link},
    volume = {12},
    copyright = {http://iopscience.iop.org/info/page/text-and-data-mining},
    issn = {1748-0221},
    shorttitle = {{MuTRiG}},
    url = {https://iopscience.iop.org/article/10.1088/1748-0221/12/01/C01043},
    doi = {10.1088/1748-0221/12/01/c01043},
    number = {01},
    journal = {Journal of Instrumentation},
    author = {Chen, H. and Briggl, K. and Eckert, P. and Harion, T. and Munwes, Y. and Shen, W. and Stankova, V. and Schultz-Coulon, H.C.},
    month = jan,
    year = {2017},
    pages = {C01043--C01043},
}

@incollection{Augustin2021JPSCP,
    series = {{JPS} {Conference} {Proceedings}},
    title = {{MuPix10}: {First} {Results} from the {Final} {Design}},
    volume = {34},
    shorttitle = {{MuPix10}},
    url = {https://journals.jps.jp/doi/10.7566/JPSCP.34.010012},
    doi = {10.7566/JPSCP.34.010012},
    number = {34},
    urldate = {2026-01-24},
    booktitle = {Proceedings of the 29th {International} {Workshop} on {Vertex} {Detectors} ({VERTEX2020})},
    publisher = {Journal of the Physical Society of Japan},
    author = {Augustin, Heiko and Berger, Niklaus and Dittmeier, Sebastian and Immig, David Maximilian and Kim, Dohun and Mandok, Lukas and Gonzalez, Annie Meneses and Menzel, Marius and Noehte, Lars Olivier Sebastian and Peri\'c, Ivan and Schmidt, Alexander and Sch\"oning, André and Vigani, Luigi and Weber, Alena and Weinläder, Benjamin},
    month = jun,
    year = {2021},
}

@book{amato_introduction_2024,
    address = {Cham},
    series = {Lecture {Notes} in {Physics}},
    title = {Introduction to {Muon} {Spin} {Spectroscopy}: {Applications} to {Solid} {State} and {Material} {Sciences}},
    volume = {961},
    isbn = {978-3-031-44958-1 978-3-031-44959-8},
    shorttitle = {Introduction to {Muon} {Spin} {Spectroscopy}},
    url = {https://link.springer.com/10.1007/978-3-031-44959-8},
    doi = {10.1007/978-3-031-44959-8},
    publisher = {Springer International Publishing},
    author = {Amato, Alex and Morenzoni, Elvezio},
    year = {2024},
}

@article{rudzki_ultra-light_2023,
    title = {An ultra-light helium cooled pixel detector for the {Mu3e} experiment},
    volume = {18},
    issn = {1748-0221},
    url = {https://doi.org/10.1088/1748-0221/18/10/C10022},
    doi = {10.1088/1748-0221/18/10/C10022},
    number = {10},
    urldate = {2026-01-24},
    journal = {Journal of Instrumentation},
    publisher = {IOP Publishing},
    author = {Rudzki, Thomas Theodor and Augustin, Heiko and Immig, David Maximilian and Kolb, Ruben and Mandok, Lukas and collaboration, on behalf of the Mu3e},
    month = oct,
    year = {2023},
    pages = {C10022},
}

@article{Stoykov2012PP,
    title = {High-{Field} $\mu${SR} {Instrument} at {PSI}: {Detector} {Solutions}},
    volume = {30},
    shorttitle = {High-{Field} μ{SR} {Instrument} at {PSI}},
    url = {https://linkinghub.elsevier.com/retrieve/pii/S187538921201214X},
    doi = {10.1016/j.phpro.2012.04.028},
    urldate = {2026-01-24},
    journal = {Physics Procedia},
    author = {Stoykov, A. and Scheuermann, R. and Sedlak, K. and Rodriguez, J. and Greuter, U. and Amato, A.},
    year = {2012},
    pages = {7--11},
}

@article{Brewer2000PBCM,
    title = {Delayed muonium formation in quartz},
    volume = {289-290},
    issn = {0921-4526},
    url = {https://www.sciencedirect.com/science/article/pii/S0921452600004270},
    doi = {10.1016/S0921-4526(00)00427-0},
    urldate = {2026-01-25},
    journal = {Physica B: Condensed Matter},
    author = {Brewer, J. H and Morris, G. D and Arseneau, D. J and Eshchenko, D. G and Storchak, V. G and Bermejo, J},
    year = {2000},
    pages = {425--427},
}

@article{Doll2025PRR,
    title = {Coherent microwave control of coupled electron-muon centers},
    volume = {7},
    url = {https://link.aps.org/doi/10.1103/zgj6-sx4h},
    doi = {10.1103/zgj6-sx4h},
    number = {3},
    journal = {Physical Review Research},
    publisher = {American Physical Society},
    author = {Doll, Andrin and Wang, Chennan and Prokscha, Thomas and Dreiser, Jan and Salman, Zaher},
    year = {2025},
    pages = {033059},
}

@book{Spieler2005,
  title = {Semiconductor Detector Systems},
  author = {Spieler, Helmuth},
  year = 2005,
  month = aug,
  publisher = {Oxford University Press},
  doi = {10.1093/acprof:oso/9780198527848.001.0001},
  isbn = {978-0-19-852784-8}
}

@misc{ToniMaxime,
  author = {Shiroka, Toni and Lamotte, Maxime},
  title = {Private communication},
  url = {https://www.psi.ch/en/smus/vms}
}

\end{document}